\documentclass[12pt]{iopart}
\usepackage{graphicx,epsfig}
\usepackage{setstack,iopams,bm}

\begin{document}

\title{ Full
counting statistics for noninteracting fermions:
Joint probability distributions}
\author {L. Inhester and K. Sch\"onhammer}
\address{Institut f\"ur Theoretische Physik, Universit\"at
  G\"ottingen, Friedrich-Hund-Platz 1, D-37077 G\"ottingen}

\date{\today}

\begin{abstract}
The 
joint probability distribution in
 the full counting statistics (FCS)
for noninteracting electrons
is discussed for 
an arbitrary number of 
 initially separate subsystems which are connected at $t=0$ and separated
 again at a later time. A simple method
to obtain the leading order long time contribution
to the logarithm of the characteristic function is presented which
simplifies earlier approaches. New explicit results for the
determinant involving the scattering matrices are found. The joint
probability distribution for the charges in {\it two} leads is
discussed for Y-junctions and dots connected to four leads.

\end{abstract}

\submitto{\JPCM}
\maketitle

\section{Introduction}

The theory of noise in quantum transport in mesoscopic
systems is a very active field of research \cite{BB,Naza}.
In addition to the first few moments of the transmitted charge 
the full probability distribution can be studied, called
{\it full counting statistics} (FCS).
This was first done 
in a publication by Levitov and Lesovik \cite{LL1} where
noninteracting fermions were treated 
using the cumulant generating function.
The system usually studied consists of a finite ``dot''-region
connected to $M$ leads which initially are separated from the dot
region and have different chemical potentials \cite{LL1,LL2,Nazarov,Bruder}.
After connecting the subsystems the time evolution of the particle
 transfer between the leads is studied.
 In order to avoid
mathematical subleties it is useful to start with
a {\it finite} total number of particles $N_{\rm tot}$ which can
be achieved using leads of finite extent. The thermodynamic limit is 
performed only at a later stage. For the lattice models

with a finite number of states at each lattice site studied in
this paper this implies also a {\it finite} number $N_H$ of the dimension of
the Hilbert space of a single fermion. 

For an initial state which is a Slater determinant with $N_{\rm tot}$
fermions the characteristic function $g(t)$
for noninteracting fermions is a
 $N_{\rm tot}\times N_{\rm tot} $ determinant. After averaging over a grand
 canonical ensemble an expression for $g(t)$ in terms of a $N_H\times
 N_H$ determinant can be derived \cite{LL2,Klich,KS}. In some
 publications this result is called ``Levitov-Lesovik formula''\cite{AI}. 
This expression is the formal starting point for the actual calculation
of the characteristic function. It consists of two steps. The first
one is
to calculate the time dependence of the one-particle 
projection operators $P_a(t)$ onto lead $a$. For finite times exact
results can be obtained numerically \cite{KS}. In the long time limit
an accurate analytical approximation can be given in terms
of scattering states after performing the thermodynamic
limit. In the second step the determinant over the one-particle Hilbert space 
has to be calculated. After the thermodynamic limit has been performed
this is an infinite dimensional determinant and mathematical care is
necessary \cite{math}. Various approaches have been proposed for
the evaluation of the determinant. Muzykantskii and Adamov \cite{MA} 
used methods from the theory of singular integral equations to
proceed for $M=2$. In the long time limit they obtained the leading term 
for $\ln {g(t)}$ (linear in $t$) by the exact solution of a matrix
Riemann-Hilbert problem. Their approach provided the first
explicit derivation of the result presented by Levitov and Lesovik
\cite{LL1}. Alternatively one can use a formal power series expansion
of the logarithm of the determinant \cite{KS}.  
The term linear in time
 can then easily be identified and the infinite dimensional determinant
can be reduced to an energy integral over the logarithm of a 
 $M\times M$ determinant. A third approach used Szeg\"o's
theorem from the theory of Toeplitz matrices to obtain the term
linear in time \cite{Bla}.

 For 
two leads ($M=2$) the contribution to $\ln {g(t)}$ linear in $t$
vanishes in the case of perfect transmission
 and subleading terms logarithmic in time
have to be considered. This was studied
analytically by  Muzykantskii and Adamov \cite{MA} by
 an approximate solution of a
more complicated Riemann-Hilbert problem as well as by extensions of
Szeg\"o's theorem \cite{Bla}. Numerical 
as well as analytical results in agreement with these findings were
presented by one of us\cite{KS}.
 
In this paper we generalize and simplify the derivation using the
formal power series expansion for $\ln {g(t)}$.
 After obtaining the
general linear in $t$ contribution the $M\times M$ determinant for the
{\it joint} probability distribution is examined in detail. The
$M \times M$  matrix of which one has to evaluate the determinant
is written as sum of the unit matrix and  a second matrix. This allows
to read off the general expression for the characteristic function
for the  joint probability distribution of two or more observed
charged transfers for arbitrary values of $M$.

Applications to Y-junctions ($M=3$) and dots with $M=4$ leads are
discussed.

\section{General formulation}

\noindent In the following we consider a system which consists
of a finite ``dot''-region described by the Hamiltonian $H_0^{\rm dot}$
and  $M$ leads with the Hamiltonians
$H_{0,a}$ with $a=1,..,M$. The leads are initially separated from the dot
region. The number of
electrons in the initial state are $N_0^{\rm dot}$ and $N_{0,a}$.
We assume the initial state $|\Phi(0)\rangle$ to be an eigenstate
of $H_0^{\rm dot}$ and the $H_{0,a}$
\begin{equation}
\label{Phi0}
|\Phi(0)\rangle=|E_i^{N_0^{\rm dot}}\rangle \otimes |E_n^{N_{0,1}}\rangle 
\otimes...\otimes |E_p^{N_{0,M}}\rangle   .
\end{equation} 
The time evolution for times greater than zero is described by
the Hamiltonian 
\begin{equation}
\label{Hamiltonian}
H=  H_0^{\rm dot}+  \sum_aH_{0,a} +\sum_{a}V_{a}\equiv H_0 +V~.
 \end{equation} 
The term $V$ which couples the leads with the dot region will be specified 
later.
The probability distribution that $Q_1$ electrons are transfered to
 lead $1$,  $Q_2$ electrons are transferred to
 lead $2$ etc. after time $t$ when the subsystems are separated
again, is given by
\begin{eqnarray}
\label{w1}
w(t,\{Q\})&=& \langle \Phi(t)|\prod_{a=1}^M\delta[Q_a-(
{\cal N}_a-N_{0,a})]|\Phi(t)\rangle  \\
&=& \frac{1}{(2\pi)^M}\int d\lambda_1...\lambda_M
e^{-i\sum_a\lambda_aQ_a}g(t,\{\lambda\})~. \nonumber
\end{eqnarray}
Here ${\cal N}_a$ ist the particle number operator of the lead $a$
and $g(t,\{\lambda\}) $ is the characteristic function. With
the particle number operators ${\cal N}_a(t) $ in the Heisenberg
picture $g$ is given by
\begin{equation}
g(t,\{\lambda\})=\langle \Phi(0)|e^{i\sum \lambda_a{\cal N}_a(t) }
 e^{-i\sum \lambda_a{\cal N}_a }  |\Phi(0)\rangle~.
\end{equation} 
The fact that the initial state is assumed to be an eigenstate of the
particle number operators was used.
This expression can be easily generalized to initial statistical
operators of the type
$\rho_0=\rho_0^{\rm dot}\otimes \rho_0^{(1)}\otimes...\otimes\rho_0^{(M)} $.
An important example are initial grand canonical subensembles with different
temperatures and chemical potentials
\begin{eqnarray}
\rho_0^{(a)}= \frac{e^{-\beta_a (H_{0,a}-\mu_a{\cal N}_a)}}
{ {\rm Tr} e^{-\beta_a(H_{0,a}-\mu_a{\cal N}_a)} }~
 \end{eqnarray}
and $\rho_0^{\rm dot} $ of the same type.
Then $\rho_0$ has the generalized canonical form $\rho_0=e^{-\bar H_0}/\bar Z_0$.
Averaging yields for the characteristic function 
\begin{equation}
\label{char}
g(t,\{\lambda\})=\langle e^{i\sum \lambda_a{\cal N}_a(t) }
 e^{-i\sum \lambda_a{\cal N}_a } \rangle~,
\end{equation} 
where $\langle...\rangle$ denotes the averaging with the statistical
operator $\rho_0$. This result is also valid for interacting fermions.\\

For noninteracting fermions the expectation value can be simplified
using Klich's trace formula\cite{Klich,KS} 
\begin{equation}
\label{Klichf}
{\rm Tr} (e^Ae^B)={\rm det}( 1 +e^{ a}e^{ b})~,
\end{equation}
where $A$ and $B$ are arbitrary one particle operators in Fock space
and $ a$ and $ b$ are the corresponding operators in the
Hilbert space of a single particle. Therefore the characteristic
function can be expressed as a determinant in the one particle Hilbert
space
\begin{eqnarray}
\label{project}
g(t,\{\lambda\})&=&{\rm det}\left [   1+\left (
    e^{i\sum_a\lambda_aP_a(t)}e^{-i\sum_a\lambda_aP_a}- 1\right
  )\bar n_0
\right]\nonumber \\
&\equiv& {\rm det}\left [  1+ b(t,\{\lambda\})\right ]~,
\end{eqnarray}
where $P_a$ is the projection operator in the Hilbert space of one
particle on the states of the $a$-th lead and $\bar n_0=(e^{\bar
  h_0}+1)^{-1}$is the Fermi operator. It is determined by
the Fermi functions describing the initial state.
 In order to obtain joint probability
distributions for arbitrary times the first step is the (numerical)
calculation of the Heisenberg operator $P_a(t)$.  

In order to study the {\it long time} behaviour it is useful 
to introduce the current operators $ j_a \equiv [P_a,h]/i$, where
$h$ is the Hamiltonian in the Hilbert space of a single particle,
and write
$P_a(t)$ as\cite{KS} 
 \begin{eqnarray}
\label{deltaP} 
P_a(t)&=&P_a(0)+\int_0^t j_a(t') dt'\equiv P_a+\delta P_a(t)~.~~
\end{eqnarray}
The operator $ b$ in Eq. (\ref{project}) can then be expressed as
\begin{equation}
  b(t,\{\lambda\})
=\sum_a (e^{i\lambda_a}-1)\delta P_a(t)e^{-i\sum_{a'}\lambda_{a'}P_{a'}}\bar
n_0~.
\end{equation}

In order to avoid reflections from the ends of the leads (far away
from the dot)
in the long time limit, the thermodynamic
limit has to be taken first. In this limit the discrete energy spectrum
of the initially occupied standing wave states $|\epsilon_j,a \rangle$ 
becomes continuous and the trace in the one-particle Hilbert space
involves an energy {\it integration} for the lead states
\begin{equation}
\label{trace1}
{\rm tr}~  b=\sum_i \langle \epsilon_i^{\rm dot}|b|\epsilon_i^{\rm dot}\rangle+
\sum_a\int d\epsilon \langle \epsilon,a|
b|\epsilon,a\rangle~,
\end{equation}
with the normalization $\langle \epsilon,a|\epsilon',a'\rangle
=\delta_{aa'}\delta(\epsilon-\epsilon') $.
Expressed with these states the projection operator onto lead $a$ reads
\begin{equation}
\label{projop}
P_a=\int d\epsilon |\epsilon,a\rangle \langle \epsilon,a| .
\end{equation}
Using ${\rm det} b=\exp({\rm tr}\ln b)$ in Eq. (\ref{project}) the logarithm
of the characteristic function is given by
\begin{equation}
\label{trace2}
\ln g(t,\{\lambda\})={\rm tr}\ln \left [  1+ b(t,\{\lambda\})\right ].
\end{equation}
 The thermodynamic limit of the trace operation is defined
in Eq. (\ref{trace1}). Mathematical subtleties concerning the
existence of the determinant in  Eq. (\ref{project}) in
the thermodynamic limit were discussed recently \cite{math}.
In the long time limit the ``dot part'' of the trace 
(see Eq. (\ref{trace1})) gives a finite contribution which will be neglected
in the following. Alternatively one can avoid the dot states
alltogether by including them in (part of) the leads\cite{KS}. 
Using $\langle \epsilon,a|\bar
n_0|\epsilon',a'\rangle=\delta(\epsilon-
\epsilon')\delta_{aa'}f_a(\epsilon)$, where the $f_a(\epsilon) $ are
the Fermi functions of the leads
the matrix elements of the operator $ b(t,\{\lambda\}) $ 
with the lead states are given by
\begin{eqnarray} 
\label{matrixel}
 \langle \epsilon_1,a_1|
 b(t,\{\lambda\})|\epsilon_2,a_2\rangle=\phantom{aaaaaaaaaaaaaaaaaa
aaaaaa} \\
\phantom{aa}
\sum_a d(\lambda_a) \int_0^t \langle \epsilon_1,a_1|e^{iht'}
j_a e^{-iht'}|\epsilon_2,a_2\rangle dt'
e^{-i\lambda_{a_2}}f_{a_2}(\epsilon_2)~,
  \nonumber
\end{eqnarray}
where $d(\lambda_a)\equiv e^{i\lambda_a}-1$.
Equations (\ref{trace2}) and (\ref{matrixel}) summarize the two tasks
in order to obtain the characteristic function in the
thermodynamic limit. In the first step the time evolution of the
matrix  element in Eq.  (\ref{matrixel}) has to be determined. In the
second step the trace in Eq.  (\ref{trace2}) has to be performed.

\section{Long time limit}

As the current operators $j_a$ only involve operators localized 
near the dot region one can approximate the time evolution in
the matrix elements in Eq. (\ref{matrixel}) in the long time limit by
$e^{-iht'}|\epsilon,a\rangle \approx
e^{-i\epsilon t'}|\epsilon,a+\rangle$,
where $ |\epsilon,a+\rangle$ is the scattering state with outgoing
boundary condition for the connected system \cite{KS}.
Then the time
integration in  Eq. (\ref{matrixel}) can be easily performed
\begin{eqnarray} 
\label{matrixelappr}
 \langle \epsilon_1,a_1|
 b(t,\{\lambda\})|\epsilon_2,a_2\rangle\approx\phantom{aaaaaaaaaaaaaaaaaa
aaaaaa} \\
\phantom{aa}
\frac{e^{i(\epsilon_1-\epsilon_2)t}-1}{i(\epsilon_1-
  \epsilon_2)}
\sum_a d(\lambda_a) \langle \epsilon_1,a_1+|
j_a|\epsilon_2,a_2+\rangle
e^{-i\lambda_{a_2}}f_{a_2}(\epsilon_2)~.
  \nonumber
\end{eqnarray}
The fact that a simple analytical expression for the matrix elements
has been derived allows systematic approximations for calculating
the trace in Eq. (\ref{trace2}). The first step is a formal power
series expansion
\begin{equation}
\label{trace3}
\ln g(t,\{\lambda\})=\sum_{n=1}^\infty\frac{(-1)^{n+1}}{n}
{\rm tr}~ b^n(t,\{\lambda\})~.
\end{equation}
In the evaluation of the trace of $ b^n$ the time dependent
prefactor of the sum in Eq. (\ref{matrixelappr}) plays a central
role.
  The product of the factors $(e^{i\epsilon_{j,l} t}-1)/\epsilon_{j,l}$
with $\epsilon_{j,l}=\epsilon_j-\epsilon_l$ 
is used to obtain a product of $n-1$ ``energy conserving'' delta functions
\begin{eqnarray} 
\label{deltafunctions}
\frac{e^{i\epsilon_{1,2}t}-1}{i\epsilon_{1,2} }
\frac{e^{i\epsilon_{2,3}t}-1}{i\epsilon_{2,3} }\dots
\frac{e^{i\epsilon_{n,1}t}-1}{i\epsilon_{n,1} } 
=\frac{\sin{(\epsilon_{1,2}t/2)}}{\epsilon_{1,2}/2 }
\frac{\sin{(\epsilon_{2,3}t/2)}}{\epsilon_{2,3}/2 } \dots
\frac{\sin{(\epsilon_{n,1}t/2)}}{\epsilon_{n,1}/2 }\nonumber \\
 \to~(2\pi)^{n-1} t\delta(\epsilon_{1}-\epsilon_{2} )
\delta(\epsilon_{2}-\epsilon_{3} )\dots
 \delta(\epsilon_{n-1}-\epsilon_{n} ) ~.
\end{eqnarray}
Therefore only {\it one} energy integration remains and the trace
in the full one particle Hilbert space can be converted to a trace
in the  $M$-dimensional space of the lead indices $a$ denoted by
$ {\rm tr}_{(M)} $
\begin{equation}
\label{trace4}
{\rm tr}~ b^n(t,\{\lambda\})\to \frac{t}{2\pi}
\int d\epsilon~ {\rm tr}_{(M)}
  c^n(\epsilon,\{\lambda\} )~,
\end{equation}
where the $M\times M$ matrix $ c(\epsilon,\{\lambda\} ) $ has 
the matrix elements
\begin{equation}
\label{ctilde}
 c_{a_1a_2}(\epsilon,\{\lambda\} )=
2\pi \sum_ad(\lambda_a)\langle \epsilon,a_1+| j_a|\epsilon,a_2+\rangle
e^{-i\lambda_{a_2}}f_{a_2}(\epsilon).
\end{equation}
In contrast to the current matrix elements  off-diagonal in
energy the diagonal elements in Eq. (\ref{ctilde}) can be
simply expressed
in terms of the scattering matrix $s_{a_1,a_2}(\epsilon)$. As shown in 
 appendix A 
\begin{equation} 
\label{currentm}
2\pi \langle \epsilon,a_1+| j_a|\epsilon,a_2+\rangle
=s^\dagger_{a_1a}(\epsilon)s^{\phantom{\dagger}}_{aa_2}(\epsilon)
 -\delta_{a_1a}\delta_{a_2a}
\end{equation} 
holds\cite{Nenciu}.
If we define the $M\times M$ matrices $e(\{\lambda\})$ and $f(\epsilon)$ as
$e_{aa'}(\{\lambda\}) \equiv e^{i\lambda_a}\delta_{aa'}$
 and $f_{aa'}(\epsilon)\equiv 
f_a(\epsilon) \delta_{aa'}  $ the matrix $c$ takes the
 form (suppressing the energy and $\lambda$ dependencies)
\begin{equation}
\label{ctilde}
c=(s^\dagger es e^\dagger-1)f\equiv \tilde c f~.
\end{equation} 
Now the relation $\ln {\rm det}(1+c)={\rm tr}\ln(1+c)$ can be used 
backwards. With Eqs. (\ref{trace3}) and (\ref{trace4}) this yields
in leading time order
\begin{equation}
\label{leading}
\ln g(t,\{\lambda\})=\frac{t}{2\pi}
\int d\epsilon~\ln {\rm det}(1+c)~.
\end{equation} 
Subleading corrections increase only logarithmically with
time\cite{MA,KS,Bla}. The leading order term was correctly given
by Levitov and Lesovik\cite{LL1} without presenting a derivation.

\section{ Evaluation of the determinant }

In order to obtain explicit results for the leading order in time 
result for $\ln g(t,\{\lambda\})$ the determinant in the integrand
of Eq. (\ref{leading}) has to be calculated
\begin{equation}
\label{detM}
D(\epsilon,\{\lambda\})\equiv {\rm det}(1+\tilde cf)
={\rm det}(1-f+s^\dagger es e^\dagger f)~.
\end{equation}
In both representations one has to calculate a determinant of a matrix
which ist the {\it sum} of a {\it diagonal} matrix and an arbitrary
matrix. The second decomposition in Eq. (\ref{detM}) is usually taken
as the starting point \cite{LL1,FSPB}. If one is interested in the
probability distribution of the transferred charge in a {\it single}
lead or the joint probability distribution of only a {\it few }
(e.g. two) of
the $Q_j$ it turns out to be preferable to use the first 
decomposition in Eq. (\ref{detM}). Then we can use
\begin{eqnarray}
\label{detformula}
{\rm det}( 1 + c )&=&
1+{\rm tr} ~ c+\sum_{\{i<j\}}{\rm det}^{(2)}c^{(2)}
+\sum_{\{i<j<k\}}{\rm det}^{(3)}c^{(3)} \nonumber \\
&+& .....+{\rm det}~c ~,
\end{eqnarray}
where e.g. ${\rm det}^{(3)}c^{(3)} $  denotes a $3\times 3$
subdeterminant of $c$ with the indices given by the summation
variables. Because of $c=\tilde c f$ the Fermi functions can be
factored out and it is sufficient to consider the subdeterminants of
$\tilde c$. The matrix elements of $\tilde c$ are given by
\begin{equation}
\tilde
c_{aa'}=\sum_bs^\dagger_{ab}(e^{i(\lambda_b-\lambda_{a'})}-1)s_{ba'}
\equiv \sum_bs^\dagger_{ab}d_{ba'}s_{ba'}~.
\end{equation}
We begin with the discussion of the 
transferred charge in a {\it single}
lead.

\subsection {Generalized Levitov-Lesovik formula}

We choose $a=1$ as the channel index of the observed charge transfer,
i.e. $\lambda_1$ is different from zero, but all $\lambda_a$ with
$a>1$ are put to zero. For $a'>1$ this implies 
\begin{equation}
\tilde c_{aa'}=\sum_bs^\dagger_{ab}(e^{i\lambda_b}-1)s_{ba'}
=s^\dagger_{a1}(e^{i\lambda_1}-1)s_{1a'}\equiv s^\dagger_{a1}d_1s_{1a'}~,
\end{equation}
while for $a'=1$ one obtains
\begin{equation}
\tilde c_{a1}=\sum_{b>1}s^\dagger_{ab}d^*_1s_{b1}=
(\delta_{a1}-s^\dagger
_{a1}s_{11})d^*_1~.
\end{equation}
Apart from the additional term $d^*_1$ in $\tilde c_{11}$ the columns
of the matrix $\tilde c$ are all proportional to $s^\dagger_{a1}$. Therefore
all subdeterminants of order larger than two on the rhs of
Eq. (\ref{detformula}) vanish.
Using $1-|s_{11}|^2=\sum_{a\ne 1}|s_{1a}|^2$ the trace term reads
\begin{equation}
{\rm tr}~c=\sum_{a\ne 1}|s_{1a}|^2(d^*_1f_1+d_1f_a)~.
\end{equation}
The only non-vanishing $2\times 2$ matrices are from the pairs $(i,j)=(1,j>1)$.
In the first column the part proportional to $s^\dagger_{a1}$ does not
contribute and with $d_1d_1^*=-(d_1+d_1^*)$ one obtains
\begin{equation} 
{\rm det}^{(2)}c^{(1a)}=-(d_1+d_1^*)|s_{1a}|^2f_1f_a~.
\end{equation}
Therefore the determinant appearing in Eq.(\ref{leading}) is given by
\begin{equation}
\label{gLL}
{\rm det}~(1+c)=1+\sum_{a\ne 1}|s_{1a}|^2\left [d_1f_a\bar f_1+
d_1^*f_1\bar f_a\right]\equiv 1+L_1~,
\end{equation}
where $\bar f_a=1-f_a$.
This is the generalized Levitov-Lesovik formula for the leading time
contribution to $\ln g(t,\lambda_1,0,...,0)$.
The derivation presented here 
simplifies an earlier one \cite{KS}.

\subsection{Joint probability distribution for two leads}

In this subsection we present the general expression for arbitrary
values of $M$ for the
characteristic function necessary to calculate the joint probability
distribution for {\it two} observed charge transfers. We evaluate
${\rm det}(1+c)$ for arbitrary values of $\lambda_1$ and $\lambda_2$
and put all $\lambda_i$ for $i>2$ to zero. For $j>2$ this implies
\begin{equation}
\label{ctilde2}
\tilde c_{ij}=\sum_ls^\dagger_{il}(e^{i\lambda_l}-1)s_{lj}
= s^\dagger_{i1}d_1s_{1j}+s^\dagger_{i2}d_2s_{2j}~.
\end{equation}  
 The third and all higher columns of the matrix $\tilde
c$ are proportional to the two column vectors $s^\dagger_{i1}$ and 
 $s^\dagger_{i2}$. Only the first two columns have an additional
 contribution. Using $d_{21}-d_1^*=d_2+d_2d_1^*$ and  $d_{12}-d_2^*=d_1+d_1d_2^*$
 they are given by
\begin{eqnarray}
\label{ctilde3}
\tilde c_{i1}&=&d_1^*\delta_{i1}-s^\dagger_{i1}d_1^*s_{11}
+s^\dagger_{i2}(d_2+d_2d_1^*)s_{21} ~, \nonumber \\
\tilde c_{i2}&=&d_2^*\delta_{i2}+s^\dagger_{i1}(d_1+d_1d_2^*)s_{12}
-s^\dagger_{i2}d_2^*s_{22}~.
\end{eqnarray}  
The structure of the matrix $\tilde c$ implies that 
only determinants up to order {\it four} on the rhs of
 Eq. (\ref{detformula}) can be different from zero.

 Of the forth order
determinants only those of the $c^{(12ij)}$ with $j>i>2$ are nonvanishing.
As new objects $2\times 2$ determinants formed by scattering matrix
elements appear
\begin{equation}
S^{(2)}_{ij}\equiv s_{1i}s_{2j}-s_{1j}s_{2i}~.
\end{equation}
For $j>i>2$ they
 contain the interference effect in the scattering of {\it two} fermions 
from the leads $i$ and $j$ to the leads $1$ and $2$.
Using again $d_id_i^*=-(d_i+d_i^*)$ one obtains
\begin{eqnarray}
\label{det4}
{\rm det}^{(4)}
c^{(12ij)}=(d_1+d_1^*)(d_2+d_2^*)|S^{(2)}_{ij}|^2f_1f_2f_if_j~.
\end{eqnarray}
The evaluation of all the determinants of order three and smaller 
is straightforward but tedious. It is therefore presented in
appendix B. With the abbreviation
\begin{equation}
\label{Bs}
B_j\equiv \frac{1}{2}(|S^{(2)}_{1j}|^2- |S^{(2)}_{2j}|^2 -|s_{2j}|^2 +|s_{1j}|^2)
\end{equation}
the general result for $D(\epsilon,\lambda_1,\lambda_2,0,...,0)=
{\rm det}(1+c)$  reads using the definition in Eq. (\ref{gLL})
\begin{eqnarray}
\label{joint}
D&=& 1+L_1+L_2 
              +[d_1d_2^*\bar f_1f_2
                  (|s_{12}|^2+\sum_{j\ne 1,2}B_jf_j )+
 (1\leftrightarrow 2)     ] \nonumber \\
              &+& \frac{1}{2} \sum_{i,j>2}|S^{(2)}_{ij}|^2
 (d_1\bar f_1 f_i+d_1^*f_1 \bar f_i)(d_2\bar f_2f_j+d_2^*f_2 \bar f_j)~. 
\end{eqnarray}
In the double sum the restriction to $i\ne j$ is included by the fact
that the $S^{(2)}_{ii}$ vanish. 

The generalization to joint probability distributions of more than two
charges is obvious but the expressions become rather lengthy. Therefore
this will not be discussed further here. Instead we next present the result
for {\it all} $\lambda_i\ne 0$ for a special form of the scattering
matrix. 

\subsection{Separable scattering matrix}

The special case of at dot consisting of a single level, called
``simple star'' is described by a separable scattering matrix
 of the form (see Appendix C)
\begin{equation}
s_{ij}=\delta_{ij}+\alpha_i\alpha_j^*u~.
\end{equation}
The unitarity of the scattering matrix implies
\begin{equation}
\label{unitary}
u+u^*+|u|^2\sum_i|\alpha_i|^2=0~.
\end{equation}
In the complex $u$-plane this is 
the equation of a circle with
radius $r_\alpha$ around $(-r_\alpha,0)$, where
$1/r_\alpha=\sum_i|\alpha_i|^2 $.
For $i\ne j$ this implies the inequality 
\begin{equation}
\label{inequ}
|s_{ij}|^2\le \frac{4|\alpha_i|^2|\alpha_j|^2 }{(\sum_i|\alpha_i|^2)^2}
\end{equation}
used later.

For the calculation of $D(\epsilon,\{\lambda\})$ it is useful
to write $ c$ as
\begin{equation}
c=(s^\dagger es-e)e^\dagger f\equiv (s^\dagger es-e)  \tilde f~.
\end{equation}
Using the unitarity relation Eq.(\ref{unitary}) and the definition
\begin{equation}
S_j\equiv \sum_l|\alpha_l|^2(e^{i\lambda_l}-e^{i\lambda_j})
\end{equation}
the matrix elements of $b$ defined by
$\alpha b\alpha^* \equiv s^\dagger es-e$, i.e.
 factoring out
the diagonal matrices $\alpha$ and $\alpha^*$, are given by
\begin{equation}
b_{ij}=(e^{i\lambda_i}-e^{i\lambda_j})u+|u|^2S_j~.
\end{equation}
It is easy to see that all determinants of submatrices of dimension three  
and larger vanish. For the $m\times m$
submatrices $i$ and $j$ take values $i_l$ with $i_1<i_2<...<i_m$. The 
determinant is unchanged if one subtracts the first row from all other ones
\begin{equation}
b_{ij}-b_{i_1j}=(e^{i\lambda_i}-e^{i\lambda_{i_1}})u~.
\end{equation}
As the second and all higher rows are proportional to each other the
subdeterminants of dimension $m\ge 3$ vanish. The $2\times 2 $
determinants are readily calculated. Using the unitarity relation
 Eq.(\ref{unitary}) the result for $D$ takes the simple form
\begin{eqnarray}
\label{separable}
D(\epsilon,\{\lambda\})&=& 1+ \sum_{i\ne j}|\alpha_i|^2|\alpha_j|^2|u|^2
d_{ij}\bar f_i f_j 
= 1+\sum_{i\ne j}|s_{ij}|^2(e^{i(\lambda_i-\lambda_j)}-1)\bar f_i
f_j \nonumber    \\
&=& 1+\sum_iL_i+\sum_{i\ne j}|s_{ij}|^2d_id_j^*\bar f_if_j 
\end{eqnarray}
For the special case of the separable scattering matrix the
interference terms discussed in subsection 4.2 and all higher ones
vanish.

\section{Shot noise}

In this section we mainly elucidate our result Eq. (\ref{joint}) which
determines the joint probability distribution for two leads 
 and
present separately 
results for $M=3$ and $M\ge 4$ in the zero temperature limit.
As we present explicit results we have to specify the
Hamiltonian introduced in Eq. (\ref{Hamiltonian}).
In the one-particle Hilbert space the leads are described as
nearest neighbor hopping chains  
\begin{equation}
h_{0a}=-\sum_{m\ge 1}(|m,a\rangle \langle m+1,a|+ {\rm H.c.})~.
\end{equation}
Two types of dots are considered.
 For the ``simple star'' the dot
hamiltonian and the coupling term are given by
\begin{equation}
h_0^{\rm dot}=V_0|0\rangle \langle 0|~,~~~v=-\sum_{a=1}^M(\tau_a|1,a\rangle
\langle 0| +{\rm H.c.})~.
\end{equation}
For the dot consisting of a ring of $M_{\rm dot}$ sites
pierced by a magnetic flux we have
\begin{equation}
h_0^{\rm dot}=\sum_{i=1}^{M_{\rm dot} }V_i|i\rangle \langle i|
+\sum_{i=1}^{M_{\rm dot}} (\tau_{i,i+1}|i\rangle \langle i+1|+{\rm H.c.} )~,
\end{equation}  
where $M_{\rm dot}+1$ corresponds to $1$. The coupling to the leads is 
assumed to be
\begin{equation}
v=-\sum_{a=1}^M(\tau_a|1,a\rangle
\langle i_a| +{\rm H.c.})~.
\end{equation}

\subsection{Y-junctions}

For $M=3$ the double sum in Eq.(\ref{joint}) is missing and  $B_3$
defined in Eq. (\ref{Bs}) takes a simpler form as
$|S^{(2)}_{23}|^2=|s_{31}|^2$ and $|S^{(2)}_{13}|^2=|s_{32}|^2$ holds \cite{LL1}.
Using the unitarity of the scattering matrix Eq. (\ref {joint})
simplifies to
\begin{eqnarray}
\label{joint3}
D= 1+L_1+L_2  
              +[d_1d_2^*\bar f_1f_2
                  (|s_{12}|^2\bar f_3 +|s_{21}|^2 f_3)+
 (1\leftrightarrow 2)     ] ~.
\end{eqnarray}
For the three leg simple star $|s_{12}|^2=|s_{21}|^2 $ holds and
Eq. (\ref{joint3}) reduces to the $M=3$
version of Eq. (\ref{separable}) with $\lambda_3=0$. At zero
temperature Eq. (\ref{joint3}) reduces to $D=1+L_1+L_2$ when lead $1$ and
$2$ have equal chemical potentials.
The correlation between the charge transfers to leads $1$ and $2$ is
still present as $\ln g$ is determined by $\ln D$.
 
The joint probability distribution shown in Fig. 1 is for
the case $\mu_3=\mu_2+\Delta \mu$. 
If the energy dependence of the scattering matrix elements can be
neglected in the energy interval $[\mu_2,\mu_3]$ the energy integration
in Eq. (\ref{leading}) can be carried out and the characteristic function $g$
is given by
\begin{equation}
g(t,\lambda_1,\lambda_2,0)=\left[ 1+|s_{13}|^2(e^{i\lambda_1}-1
)  +|s_{23}|^2(e^{i\lambda_2}-1)   \right]^{N_t}~,
\end{equation}
where $N_t=t\Delta \mu/(2\pi)$. For integer values of $N_t$ the
$\lambda$-integrations in Eq. (\ref{w1}) can easily be performed
and the $p_{nm}(t)$ in
\begin{equation}
w(t,Q_1,Q_2)= \sum_{n,m}p_{nm}(t)\delta(Q_1-n)\delta(Q_2-m)
\end{equation}
are given by 
\begin{equation}  
\label{pnm}
p_{nm}(t)={N\choose n+m} {n+m \choose m} T_1^nT_2^m(1-T_1-T_2)^{N-(n+m)}, 
\end{equation}
where the $T_i=|s_{i3}|^2$ are the transmission probabilities from
lead $3$ to leads $1$ and $2$.

 For the ``simple star''
 with $V_0=0,~ \tau_1= \tau_3= 1$
 the transmission probabilities in the middle of the band are given by (see
Appendix C)
\begin{equation}  
T_1=\frac{4}{(2+|\tau_2|^2)^2}~,~~~~T_2=\frac{4|\tau_2|^2 }{(2+|\tau_2|^2)^2}~.
\end{equation}

For $\tau_2=0$ the connected system corresponds to an infinite ideal
chain with perfect transmission $T_1=1$. If the coupling to lead $2$
is switched on $T_1$ decreases and equals $T_2=4/9$ for equal
coupling $\tau_2=1$.

\vspace{0.1cm}
\begin{figure}[hbt]
\begin{center}
\vspace{0.5cm}
\leavevmode
\epsfxsize10.5cm
\epsffile{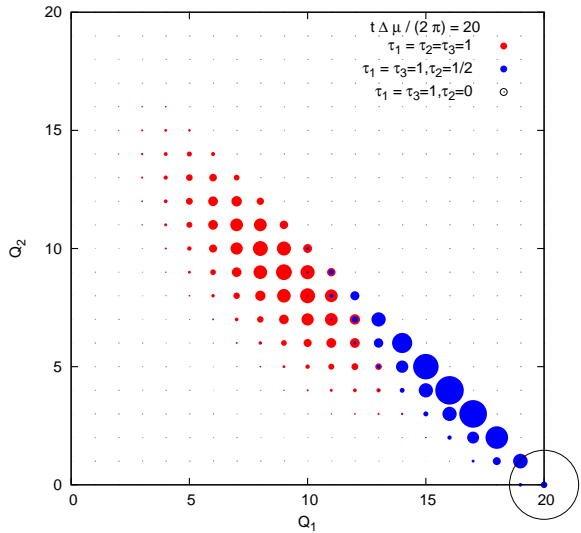}
\caption {Joint probability distribution for a three leg simple star
  for $N_t=20$ and three different coupling strengths $\tau_2$ to lead $2$
  using the leading order time approximation for $\ln g$. The area of
the dots is proportional to the $p_{nm}$ in Eq. (\ref{pnm}). 
 }
\label{Levitov07}
\end{center}
\vspace{0.0cm}
\end{figure}

 Fig. 1 shows the probabilities $p_{nm}(t)$ for $N_t=20$
and three values of $\tau_2$.  
For $\tau_2=0$ the open circle shows the ``vanishing shot noise''
which is an artefact of the leading order in $t$ approximation for
$\ln g$. The neglected logarithmic corrections convert the single
delta function to a Gaussian with a width proportional to
$(\ln{N_t})^{1/2}$ for $N_t\gg 1$ \cite{MA,KS,Bla}. For
a weak coupling to lead $2$ ($\tau_2=0.5$, dark dots) 
the anticorrelation between the transported charges to leads $1$ and
$2$ is clearly visible. It was 
discussed previously on the level of moments \cite{BB}. The light
dots correspond to the case of equal couplings to the star.

\subsection{Systems with $M\ge 4$ leads}

For Y-junctions there are no interference effects in two particle scattering
processes. The double sum term on the rhs of Eq. (\ref{joint}) only
contributes for $M\ge 4$. To simplify the discussion we here discuss
only the special case $\mu_1=\mu_2 \equiv \mu_R$, i.e. 
$f_1=f_2\equiv f_R$ at zero
 temperature
which implies $f_R^2=f_R$. Then Eq. (\ref{joint}) simplifies to 
\begin{eqnarray}
\label{jointspecial}
D= 1+L_1+L_2  
              + \frac{1}{2} \sum_{i,j>2}|S^{(2)}_{ij}|^2
 (d_1 d_2 \bar f_Rf_if_j +d_1^* d_2^*f_R \bar f_i \bar f_j)~. 
\end{eqnarray}
Already for $M=4$ there are various possibilities for the chemical potentials
of the leads $3$ and $4$. The simplest one is to assume them to be 
equal. For arbitrary $M\ge 4$ and $\mu_i=\mu_L=\mu_R+\Delta\mu$
 for all $i> 2$
Eq. (\ref{jointspecial}) further simplifies for the energy interval
$\mu_R<\epsilon <\mu_L$ to 
\begin{eqnarray}
\label{jointspecial2}
D= 1+\sum_{i=1,2}d_i\sum_{j>2}|s_{ij}|^2  
              +d_1d_2 \sum_{j>i>2}|S^{(2)}_{ij}|^2~.
\end{eqnarray}
The interference terms in the double sum also occur in the
distribution of the total charge $Q_1+Q_2$ in the leads with chemical
potential $\mu_R$ which can be obtained by putting
$\lambda_1=\lambda_2=\lambda$. As the prefactor
$d^2=(e^{i\lambda}-1)^2$ has no contribution linear in $\lambda$ the
interference terms only enter the cumulants $\kappa_i$ with $i \ge 2$. 
When the energy dependence of the scattering matrix for
$\mu_R<\epsilon <\mu_L$ is neglected the probability distribution 
for integer $N_t=t\Delta \mu /(2\pi)\gg 1$ is given by
\begin{equation}  
w^{(S)}(t,Q_1+Q_2)= \sum_{n\ge 0}p_n^{(S)}(t)\delta (Q_1+Q_2-n )
\end{equation} 
with

\begin{equation}  
p_n^{(S)}=\sum_{l=l_0}^{[n/2]}{N_t \choose n-l}{n-l \choose
  l}(1-A+B)^{N_t-n+l}B^{l}(A-2B)^{n-2l},
\end{equation}  
where $l_0={\rm max}(0,n-N_t)$ and $A$ and $B$ are given by
\begin{equation} 
A=\sum_{i=1}^2\sum_{j>2}|s_{ij}|^2~,~~~~~B=\sum_{j>i>2}|S^{(2)}_{ij}|^2~.
\end{equation}
In Fig. 2 we show results for the $p_n^{(S)} $ for a symmetric ring
dot with $\tau_a=0.5,~V_a=0$ and $ \tau_{i,i+1}=e^{i\Phi/M_d}$, where
$\Phi$ is the magnetic flux through the ring. We choose $M=M_d=4$
with each ring site connected to one lead. 
 The scattering matrix elements needed  are presented
in Appendix C . In addition to the probability distribution
$w^{(S)}$ of the sum $Q_1+Q_2$ (filled symbols) we show the 
\vspace{0.1cm}
\begin{figure}[hbt]
\begin{center}
\vspace{0.5cm}
\leavevmode
\epsfxsize9.5cm
\epsffile{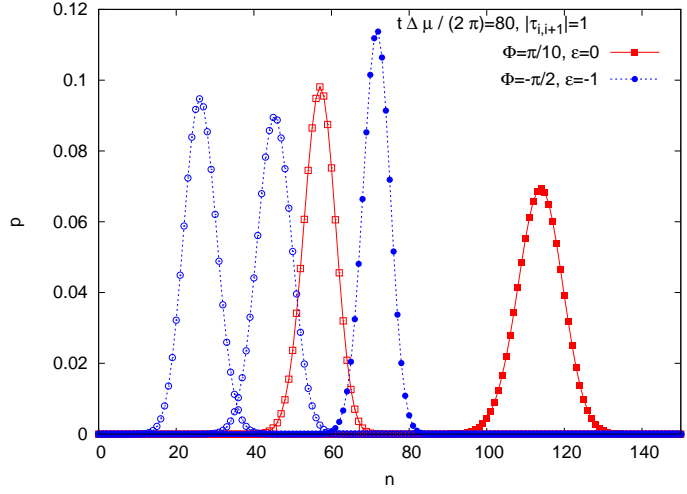}
\caption { Probability distributions for a four site ring pierced by
a magnetic flux $\Phi$. The full symbols present the $p^{(S)}_n$ of the 
probability distribution of the sum $Q=Q_1+Q_2$. The open symbols
show the probability distribution for the charge transfer to a single
lead. The lines are a guide to the eye.
Because of the magnetic flux the distributions for lead $1$ and $2$
(the two curves with open circles)
differ except in the particle-hole symmetric case (open squares).}
\label{Levitov07}
\end{center}
\vspace{0.0cm}
\end{figure}
 probability distributions for the charge transfer to the single
leads $1$ and $2$ (open symbols) which follows from Eq.(\ref{gLL}).
The circles correspond to the case when the chemical potentials
are close to the center of the band. Because
$|s_{12}|^2=|s_{14}|^2$ holds for $\epsilon=0$ for all values of
$\Phi$ the probability distributions for the charge transfer to
leads $1$ and $2$ are identical. In the generic case they are
different for $\Phi\ne 0 (\rm{mod} 2\pi)$ as shown for $\epsilon=-1$
(circles).  
For the parameters chosen it is clearly 
visible that the sum of the widths of these single charge probability
distributions is smaller than the width of $w^{(S)}$. This is
another manifestation of the
anticorrelation effect \cite{BB} mentioned in the discussion of
Fig. 1. 

The joint probability distribution $w(t,Q_1,Q_2)$ corresponding to the
parameter values used in Fig. 2 is shown in Fig. 3. In contrast to
Fig. 1 where the asymmetry in the (dark dot) distribution resulted
from asymmetric couplings, the asymmetry of the $\epsilon =-1$
distribution
in Fig. 3 is due to the magnetic flux.\\

\vspace{0.1cm}
\begin{figure}[hbt]
\begin{center}

\vspace{0.5cm}
\leavevmode
\epsfxsize11.5cm
\epsffile{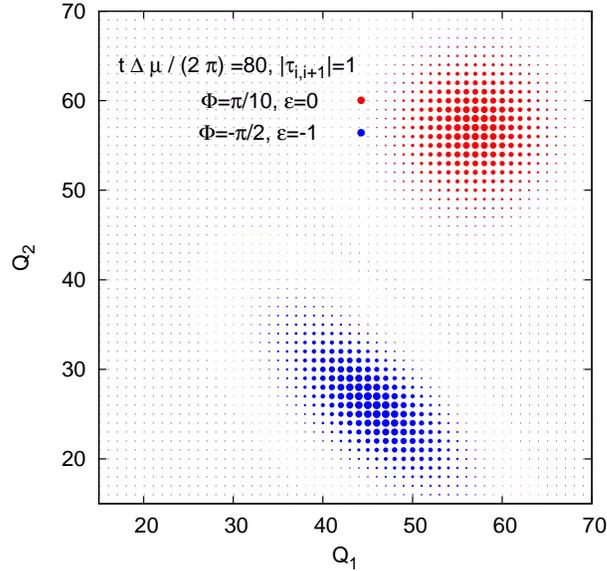}
\caption { Joint probability distributions for a four site ring for
  the paramters used in Fig. 2  }
\label{Levitov07}
\end{center}
\vspace{0.0cm}
\end{figure}

As a simple application of Eq. (\ref{separable}) 
for the separable scattering matrix we consider the
distribution of the total charge $Q_R=Q_1+...+Q_{M_R}$ transferred
to the first $M_R$ channels which are all assumed to have the chemical
potential $\mu_R$. For the case where the other $M_L=M-M_R$
 channels have chemical
potential $\mu_L=\mu_R+\Delta \mu$  Eq. (\ref{separable}) reduces
to the $M=2$ Levitov-Lesovik formula Eq. (\ref{gLL}) 
with the effective transmission
probability
\begin{equation}
T_{\rm eff}=\sum_{i=1}^{M_R}\sum_{j=M_R+1}^M|s_{ij}|^2 \le 1~.
\end{equation}
The inequality follows from Eq. (\ref{inequ}) or the explicit result
for the simple star in appendix C.

\section{Summary}

A simple derivation of the leading time order result for the logarithm
of the characteristic function which determines the full counting
statistics for systems of noninteracting fermions was presented.
The energy dependent determinant involving the scattering matrix
of the $M$-lead system was simplified analytically for three cases.
For the distribution of charge transfer to a single lead only the
absolute values of the scattering matrix elements enter. For the
joint probability distribution of the charges in two leads
interference effects in the scattering of pairs of particles
become important for $M\ge 4$.
For a separable scattering matrix which describes a simple star like
geometry the $M\times M$ determinant was evaluated for
joint probability distributions for an arbitrary number of observed
charges. All interference terms vanish and  only the
absolute values of the scattering matrix elements enter as in the
generalized Levitov-Lesovik formula.

 Explicit results for the probability
distributions were presented for the simple star
and a ring pierced by a magnetic flux at zero temperature. 
Various manifestions of the anticorrelation effects in the charge
transfer to the observed leads were shown.

Finite temperature effects will be published in a forthcoming publication.

\begin{appendix}
\section{Proof of Eq. (\ref{currentm}) }
In this appendix we present a proof of the relation for current
 matrix elements with equal
energies \cite{Nenciu}.

The scattering states $|\epsilon,a+\rangle$ introduced in section 3
obey the Lippmann-Schwinger equation\cite{Taylor}
\begin{equation}
|\epsilon,a+\rangle= |\epsilon,a\rangle+g_0(\epsilon+i0)v|\epsilon,a+\rangle~,
\end{equation}
where $v$ ist the operator which describes the
connection of the leads
with the dot region and $g_0(z)=(z-h_0)^{-1}$ ist the unperturbed
resolvent.

 As the projection operators $P_a$ commute with $h_0$ 
the current matrix elements are given by
\begin{equation}
\langle \epsilon,a_1+| j_a|\epsilon,a_2+\rangle=\frac{1}{i}
\langle \epsilon,a_1+|[P_a,v]|\epsilon,a_2+\rangle~.
\end{equation}
Using Eq.(\ref{projop}) for $P_a$ the first term of the commutator reads
\begin{equation}
\langle \epsilon,a_1+|P_av|\epsilon,a_2+\rangle
=\int d \epsilon' \langle \epsilon,a_1+|\epsilon',a\rangle \langle
 \epsilon',a|v|\epsilon,a_2+\rangle.
\end{equation}
With help of the  Lippmann-Schwinger equation
the overlaps of the unperturbed states with the scattering states can
be written as
\begin{equation}
\langle \epsilon,a_1+|\epsilon',a\rangle=\delta(\epsilon-\epsilon')\delta_{a_1a}
+\frac{\langle \epsilon,a_1+|v|\epsilon',a\rangle }{\epsilon-\epsilon'-i0}~.
\end{equation}
This yields
\begin{eqnarray}
\label{first}
\langle \epsilon,a_1+|P_av|\epsilon,a_2+\rangle
&=& \langle \epsilon,a|v|\epsilon,a_2+\rangle \delta_{a_1a} \\
&+& \int d \epsilon' \frac{\langle \epsilon,a_1+|v|\epsilon',a\rangle
\langle \epsilon',a|v|\epsilon,a_2+\rangle}{\epsilon-\epsilon'-i0} ~.\nonumber
\end{eqnarray}
The matrix element of $vP_a$ can be calculated correspondingly. In the
second term in Eq. (\ref{first}) the 
$-i0$ in the denominator
is replaced by $i0$.  This leads to a delta function
for the commutator and the energy integration can be carried out.
 One obtains
\begin{eqnarray}
\label{current}
\langle \epsilon,a_1+|  j_a|\epsilon,a_2+\rangle &=&
i \langle \epsilon,a_1+|v|\epsilon,a\rangle \delta_{a_2a}
-i \langle \epsilon,a|v|\epsilon,a_2+\rangle \delta_{a_1a} \nonumber \\
&+& 2\pi \langle \epsilon,a_1+|v|\epsilon,a\rangle
 \langle \epsilon,a|v|\epsilon,a_2+\rangle.
\end{eqnarray}
With the definition of the scattering matrix $s_{aa'}(\epsilon)$ for
 potential scattering
 \cite{Taylor}
\begin{equation}
\label{smatrix}
s_{aa'}(\epsilon)=
\delta_{aa'}-2\pi i\langle \epsilon,a|v|\epsilon,a'+\rangle 
\equiv \delta_{aa'}-2\pi i t_{aa'}(\epsilon)
\end{equation}
the  validity of
Eq. (\ref{currentm}) directly follows.

The unitarity of the scattering matrix can be explicitely confirmed 
for the multi-lead system \cite{KS}.
\section{Derivation of Eq. (\ref{joint})}

In this appendix we discuss the $m\times m $ subdeterminants of the
matrix $\tilde c$ defined in Eq. (\ref{ctilde}) for $\lambda_i=0$
for all $i\ge 3$. They are characterized by the index set $i_n\in [1,M]$
with $i_1<i_2<...<i_m$. The corresponding submatrices are denoted by
$\tilde c^{(i_1,i_2,...,i_m)}$.  The matrix elements
of these submatrices can be directly read off the matrix elements of the full
$M\times M$ matrix $\tilde c$
\begin{equation}
\tilde c_{ij}=\delta_{i1}\delta_{1j}d_1^*+\delta_{i2}\delta_{2j}d_2^*
+s_{i1}^\dagger \alpha_j+s_{i2}^\dagger \beta_j~,
\end{equation}
where the $\alpha_j$ and $\beta_j$ are given in
Eqs. (\ref{ctilde2}) and (\ref{ctilde3}). Like the full matrix $\tilde
c$ all columns of the submatrices are linear combinations of at most 
the four column vectors $\delta_{i1}, \delta_{i2},s_{i1}^\dagger $  
and $s_{i2}^\dagger$. Therefore all subdeterminants with $m>4$ vanish. 
In the following we discuss the cases $m=4,3,2$ separately.

$m=4$: 

Only for the index set $(1,2,i,j)$ with $2<i<j$ all four
different column vectors appear. Therefore all other subdeterminants
vanish. For the determinant of $\tilde c^{(1,2,i,j)}$ the following
formula is used. Let $a\equiv (s_{i1}^\dagger,s_{j1}^\dagger)^T$ 
and $b\equiv (s_{i2}^\dagger,s_{j2}^\dagger)^T$
two column vectors. Then the $2\times 2$ determinant of
linear combinations of the two vectors is given by
\begin{eqnarray}
\label{lemma}
{\rm det}(a\alpha_i+b\beta_i,a\alpha_j+b\beta_j)
=(\alpha_i\beta_j- \alpha_j\beta_i){\rm det}(a,b) \nonumber \\
= (\alpha_i\beta_j- \alpha_j\beta_i)(a_ib_j-a_jb_i)~.
\end{eqnarray}
The use of this relation and Eqs. (\ref{ctilde2}) and (\ref{ctilde3})
 yield Eq. (\ref{det4}).

$m=3$:

All subdeterminants of $\tilde c^{ijk}$ with $2<i<j<k$ vanish. For
$i=1$ or $i=2$ and $2<j<k$ the calculation proceeds using
Eq.(\ref{lemma})
\begin{equation}
{\rm det}\tilde
c^{(2jk)}=d_1d_2d_2^*|s_{j1}s_{k2}-s_{k1}s_{j2}|^2=-d_1(d_2+d_2^*)
|S^{(2)}_{jk}|^2~.
\end{equation} 
In $\tilde c^{(12i)}$ with $i>2$ the first two columns $l=1,2$ are of the
type $\alpha_la+\beta_l b +d_l^*e_l$  were $e_l$ is the $l$'th unit
column vector while in the third column $\alpha_ia+\beta_i b$ this
additional contribution is missing. Therefore the determinants
${\rm det}(e_l,a,b)$ and ${\rm det}(e_1,e_2,a(b))$ have to be
evaluated. This leads to

\begin{equation}
{\rm det}\tilde
c^{(12i)}=(d_1+d_1^*)d_2^*(|S^{(2)}_{2i}|^2-|s_{1i}|^2)
+(1\leftrightarrow 2) ~.
\end{equation}
For $m=2$ arbitrary combinations of $i<j$ in ${\rm det}\tilde
c^{(ij)}$ contribute. The evaluation is straightforward using
Eq. (\ref{lemma}).

 In order to obtain the form presented in
Eq.(\ref{joint}) the following ``sum rules'' for the $|S^{(2)}_{ij}|^2 $
were used which follow from the unitarity of the scattering matrix
\begin{eqnarray}
\sum_{i>2}|S^{(2)}_{ij}|^2&=&|s_{1j}|^2+|s_{2j}|^2-|S^{(2)}_{1j}|^2
-|S^{(2)}_{2j}|^2 ~,\nonumber \\
\sum_{j>i>2}|S^{(2)}_{ij}|^2&=&1+|S^{(2)}_{12}|^2-|s_{11}|^2-|s_{22}|^2
-|s_{12}|^2-|s_{21}|^2~. \nonumber
\end{eqnarray}

\section{ Results for the scattering matrices}

In this appendix the scattering matrices for the two dot models
presented in section V are calculated. The $t$-matrix defined 
in Eq. (\ref{smatrix}) can be expressed via the resolvent
operator $g(z)=(z-h)^{-1}$ as \cite{Taylor}
\begin{equation}
t_{aa'}(\epsilon)=\langle \epsilon,a|v|\epsilon,a'\rangle+
\langle \epsilon,a|vg(\epsilon+i0)v|\epsilon,a'\rangle~.
\end{equation}
For both dot models discussed in section 5 the first term on the
rhs vanishes. With $|\langle \epsilon,a|1,a\rangle |^2=
\sqrt{1-(\epsilon/2)^2}/\pi$ this yields
\begin{equation}
\label{sexplicit}
s_{aa'}(\epsilon)=\delta_{aa'}-2i\tau_a\tau_{a'}
\sqrt{1-(\epsilon/2)^2}
   \langle i_a|g(\epsilon+i0)|i_{a'}\rangle.
\end{equation}
For the simple dot all $|i_a\rangle$ are given by $|0\rangle$ and the
scattering matrix is of the separable form discussed in section 4.3.
 
Using the projection onto the states on the ring the full resolvent 
matrix elements in Eq. (\ref{sexplicit}) can be written as \cite{KS}
\begin{eqnarray}
\label{dotinversion}
 \langle i_a|g(z)|i_{a'}\rangle = \langle i_a|[z-h^{\rm dot}_0-
\gamma(z)]^{-1}|i_{a'}
\rangle
\end{eqnarray} 
with
\begin{eqnarray}
\label{Gamma}
\gamma(z)=g_b^0(z)\sum_{a=1}^M|\tau_a|^2|i_a\rangle \langle i_a|~,
\end{eqnarray} 
where $h^{\rm dot}_0$ is the one particle Hamiltonian
of the ring and $ g_b^0(z)\ $ is the diagonal element of the resolvent
of the semi-infinite chain at the boundary. For $z=\epsilon\pm i0$ it is
given by
\begin{eqnarray}
 g_b^0(\epsilon\pm i0)=(\epsilon\mp i\sqrt{4-\epsilon^2})/2~.
\end{eqnarray} 
The matrix inversion in Eq. (\ref{dotinversion}) is trivial for the
simple star because $M_{\rm dot}=1$
\begin{equation}
\langle 0|g(z)|0\rangle=\frac{1}{z-V_0-\sum_a|\tau_a|^2 g_b^0(z) }~.
\end{equation}
 For a symmetric ring with
$M=M_{\rm dot}$,
$\tau_a=\tau,~V_a=V_0$, and $\tau_{i,i+1}=\tau_{\rm dot}e^{i\Phi/M}$ the matrix
inversion is straightforward using plane wave states on the dot\cite{Xavier}
 \begin{eqnarray}
 \langle i_m|g(z)|i_n\rangle
=\frac{1}{M}\sum_{j=1}^M\frac{e^{ik_j(m-n)}}{z-V_0-|\tau|^2 
 g_b^0(z)+2\tau_{\rm dot}\cos{(k_j+\Phi/M)}} \nonumber~,
\end{eqnarray} 
where $k_j=2\pi j/M$.

\end{appendix}

\end{document}